\documentclass[article]{aa}
\usepackage{graphicx}
\usepackage{ulem}
\usepackage{color}
\usepackage{amsmath,amssymb}
\usepackage[varg]{txfonts}

\def\mso{\,\mathrm{M}_\odot}
\def\rso{\,{\rm R}_\odot}

\def\zso{\,{\rm Z}_\odot}
\def\simle{\mathrel{\hbox{\rlap{\hbox{\lower4pt\hbox{$\sim$}}}\hbox{$<$}}}}
\def\simgr{\mathrel{\hbox{\rlap{\hbox{\lower4pt\hbox{$\sim$}}}\hbox{$>$}}}}

\begin{document}
	
\title{A new route towards merging massive black holes} 
	
\author{Pablo Marchant\inst{1}\thanks{Email: pablo@astro.uni-bonn.de}
\and Norbert Langer\inst{1} 
\and Philipp Podsiadlowski\inst{2,1}
\and Thomas M. Tauris\inst{1,3}
\and Takashi J. Moriya\inst{1}} 
		
\institute{Argelander-Institut f\"ur Astronomie, Universit\"at Bonn, Auf dem H\"ugel 71, 53121 Bonn, Germany
\and
Department of Astrophysics, University of Oxford, Oxford OX1 3RH, UK
\and
Max-Planck-Institut f\"ur Radioastronomie, Auf dem H\"ugel 69, 53121 Bonn, Germany}
	
\date{}

\abstract {With recent advances in gravitational-wave astronomy, the
  direct detection of gravitational waves from the merger of two
  stellar-mass compact objects has become a realistic prospect.
  Evolutionary scenarios towards mergers of various double compact
  objects generally invoke so-called common-envelope evolution which
  is poorly understood, leading to large uncertainties in the
  predicted merger rates. Here we explore, as an alternative, the
  scenario of massive overcontact binary (MOB) evolution, which
  involves two very massive stars in a very tight binary which remain
  fully mixed due to their tidally induced high spin. While many of
  these systems merge early-on, we find large numbers of MOBs which
  swap mass several times but survive as a close binary until the
  stars collapse.  The simplicity of the MOB scenario allows us to use
  the efficient, public stellar-evolution code MESA for exploring it
  systematically by means of detailed numerical calculations.  We find
  that, at low metallicity, MOBs produce double-black-hole (BH+BH)
  systems that will merge within a Hubble time with mass-ratios close
  to one, in two mass ranges, $\sim25\dots 60\mso$ and $\simgr
  130\mso$, with pair-instability supernovae (PISNe) being produced at
  intermediate masses. Our models are also able to reproduce
  counterparts of various stages in the MOB scenario in the local
  Universe, providing direct support for the scenario.  We map the
  initial binary parameter space that produces BH+BH mergers,
  determine the expected chirp mass distribution, merger times, the
  expected Kerr parameters and predict event rates.  We typically find
  that for $Z\simle Z_\odot/10$, there is one BH+BH merger event for
  $\sim 1,000$ core-collapse supernovae. The advanced LIGO (aLIGO)
  detection rate is more uncertain and depends on the metallicity
  evolution. Deriving upper and lower limits from a local and a global
  approximation for the metallicity distribution of massive stars, we
  estimate aLIGO detection rates (at the aLIGO design limit) of $\sim
  19-550\,$yr$^{-1}$ for BH-BH mergers below the PISN gap and of $\sim
  2.1-370\,$yr$^{-1}$ above the PISN gap.  Even with conservative
  assumptions, we find that aLIGO should soon detect BH+BH mergers
  from the MOB scenario and that these could be the dominant source
  for aLIGO detections.  }
	
\keywords{-- stars: binaries (including multiple): close -- stars: rotation
-- stars: black holes -- stars: massive -- gravitational waves}

\maketitle 

\section{Introduction}
\label{sect:intro}

The existence of stellar binaries which consist of two stellar
remnants so close that they can merge within the Hubble time have been
known for a long time.  Merging double-white-dwarf systems are thought
to provide one potential channel to produce so called Type\,Ia
supernovae, which are the main producers of iron and through which the
accelerated expansion of the Universe was discovered.
Also, various double-neutron-star systems have been found, most
importantly the Hulse-Taylor system {\citep{HulseTaylor1975,Weisberg2010}} and more recently the double pulsar PSR J0737-3039
{\citep{Burgay2003,Kramer2006}}, whose orbital decay has
confirmed Einstein's gravitational-wave prediction immaculately.
Double-black-hole binaries have not yet been detected, which is
obviously difficult since they do not emit electromagnetic radiation.
However, as they are the most massive of the double compact binaries,
their gravitational-wave radiation would be much stronger than that of
white-dwarf or neutron-star systems.

The evolution of a binary system from the initial stage of two
orbiting main-sequence stars to the double compact binary stage is
believed to require, in most scenarios, at least one so called
common-envelope phase. This is essential as stars tend to expand dramatically after their main-sequence evolution, and as an early merging of
the binary is to be avoided, a large orbital separation is required to
accommodate this. The common-envelope stage is then necessary to
shrink the system to a compact final configuration. Whereas the
physics of this process is not yet well understood \citep{Ivanova13},
some observational constraints exist for low-mass binaries (e.g., \citealt{Han2003,Zorotovic2011}).
For massive binaries, observational evidence is much scarcer, and the
common-envelope process is much less well understood theoretically
\citep{Taam2000}.  For double-neutron-star systems, the
increased number of observed systems is beginning to provide some
constraints {\citep{Kalogera2004}}.  For double-black-hole systems, however, the
uncertainties are so large that the predicted rates of black-hole
binaries from the common-envelope channel are uncertain by several
orders of magnitude \citep{Abadie10}.

A completely different route towards double compact binaries
is explored in this paper, which does not involve a common-envelope phase
and may work only for very massive stars: it involves the chemically
homogeneous evolution of rapidly rotating stars in tidally locked binaries.

Usually the stabilizing effects of entropy and composition gradients
prevent the efficient mixing of gas once composition gradients have
been established. However, the more massive a star is, the less this
is the case due to the increased importance of radiation pressure.
It has been found that mixing induced by rapid rotation can, in
principle, keep massive stars chemically homogeneous
throughout core hydrogen burning \citep{Maeder87,Langer92,Heger00}. 
Detailed studies of this type of
evolution through large grids of stellar models
\citep{Yoon05,Woosley06,Brott11,Kohler15,Szecsi15} have shown that 
this works only at low
metallicity where strong angular-momentum loss due to stellar winds
can be avoided such that the stars remain fully mixed until core
hydrogen exhaustion.  Since, in this case, rapidly spinning iron cores
are produced at the end of the stars' lives, such single-star models
have been suggested as progenitors of long-duration gamma-ray bursts
(LGRBs) \citep{Woosley06,Yoon06}.

Chemically homogeneously evolving stars avoid the strong post-main
sequence expansion as they do not maintain a massive hydrogen-rich 
envelope. \citet{Mink09} therefore suggested that massive
close binaries, where rapid rotation and thus chemically homogeneous
evolution can be enforced through the tidal interaction of
both stars, could evolve towards black holes without ever
encountering contact or mass transfer (see also \citet{MandelMink16, Song16}).

In this paper, we explore the evolution of close binaries with
component masses above $\sim 20\mso$ by computing large grids of
detailed binary evolution models.  For this purpose, we use the
publicly available code MESA, which we extended to allow the inclusion
of contact binaries with mass-ratios close to one (Sect.\,2). Against
our initial expectation, we find that contact-free evolution occurs
only very rarely.  Instead, when computing the evolution of massive
overcontact binaries (MOBs), we find many systems which avoid merging
during core hydrogen burning.
We compute the final configurations of these binaries in Sect.\,3,
including the black-hole masses, separations, and their mass-ratios.
In Sect.\,4 we discuss the predicted black-hole Kerr parameters,
potential explosive mass loss and momentum kicks, and the connection
of the MOB scenario with long-duration gamma-ray bursts (GRBs) and
pair-instability supernovae (PISNe).  We discuss event rates and
potential LIGO detection rates in Sect.\,5, before giving our
conclusions in Sect.\,6.

\section{Methods}
\label{sect:methods}

In this paper, we provide the first detailed binary stellar evolution models
which are followed until the double-black-hole stage.  
To obtain those, we apply the
MESA code \citep{Paxton15,Paxton13,Paxton11}, which now includes all the physics required
for such calculations, in particular, tidal interactions and
differential rotation. We have computed $\sim 2000$ detailed
binary-evolution sequences in 6 model grids for different initial
metallicities and mass-ratios, thereby
achieving a complete coverage of the relevant parameter space.  Our
model calculations include the overcontact phase which occurs in
the closest simulated binaries, which
constitutes the main channel for providing 
massive close black-hole binaries.

\subsection{Physics implemented in MESA and initial parameters}
\label{sect:mesa}

To model the evolution of our systems, we have used the {\it binary}
module in version r8118 of the MESA code.\footnote{The necessary input files to reproduce
	our results with this MESA version will be provided at \url{http://mesastar.org/}.}
Opacities are calculated
using CO-enhanced opacity tables from the OPAL project
\citep{Iglesias96}, computed using solar-scaled abundances based on
\cite{Grevesse96}. Convection is modelled using the standard
mixing-length theory \citep{BohmVitense58} with a mixing length
parameter $\alpha=1.5$, adopting the Ledoux criterion. Semiconvection
is modelled according to \cite{Langer83} with an efficiency
parameter $\alpha_{\mathrm{sc}}=1.0$. We include convective core
overshooting during core hydrogen burning following \cite{Brott11}.
The effect of the centrifugal force is implemented as in
\cite{Heger00}. Composition- and angular-momentum transport due to
rotation includes the effects of Eddington-Sweet circulation, secular
and dynamical shear instabilities and the GSF instability, with an
efficiency factor $f_c=1/30$. This corresponds to the
calibrations of the mixing efficiency in stellar models based on the
VLT FLAMES Survey of Massive Stars (\citealt{Brott11}, and references
therein).  We include the effect of magnetic fields on the transport
of angular momentum as in \cite{Petrovic05}. Tidal effects are
implemented as in \cite{Hurley+02} and \cite{Detmers08} for the
case of stars with a radiative envelope. As we are not interested in
following the nucleosynthesis in detail, we use the simple networks
provided with MESA \texttt{basic.net} for H and He burning,
\texttt{co\_burn.net} for C and O burning and \texttt{approx21.net}
for later phases.

Our implementation of stellar winds follows that of \cite{Yoon06},
with mass loss rates for hydrogen-rich stars (with a surface helium abundance $Y_\mathrm{s}<0.4$)
computed as in \cite{Vink01}, while for hydrogen-poor stars
($Y_{\mathrm{s}}>0.7$) we use the recipe of \cite{Hamann95}
multiplied by a factor of one tenths. In the range
$0.4<Y_{\mathrm{s}}<0.7$, the rate is interpolated between the two. For
both rates we use a metallicity scaling of
$(Z/Z_\odot)^{0.85}$. We also include the enhancement of winds through
rotation as in \cite{Heger00}, and, when the rotation rate exceeds a
given threshold $\Omega/\Omega_\mathrm{crit}>0.98$, we implicitly compute
the mass-loss rate required for the rotation rate to remain
just below this value.

Whenever one component in the system attempts to overflow its Roche
lobe, we implicitly compute the mass-transfer rate necessary for it to
remain just within the Roche volume (computed as in \citealt{Eggleton83}). The treatment of mass transfer in over-contact systems is
described in the following section.

We consider four different metallicities, $Z_\odot/4$, $Z_\odot/10$,
$Z_\odot/20$ and $Z_\odot/50$, with $Z_\odot=0.017$ as in \cite{Grevesse96},
and the helium abundance is set in such a way that it
increases linearly from its primordial value $Y=0.2477$ \citep{Peimbert07}
at $Z=0$ to $Y=0.28$ at $Z=Z_\odot$. For all metallicities,
we compute grids for mass-ratios $q_\mathrm{i}=M_2/M_1=1$; for $Z_\odot/50$, we also compute grids at
$q_\mathrm{i}=0.9,0.8$. The initial orbital periods are chosen from the range
($P_{\rm i}=0.5-3.0$) days, with an interval of $0.1$ days, while the initial
primary masses are taken from the range ($\log\;M_1/M_\odot=1.4-2.7$) in
intervals of $0.1$ dex.

\subsection{The computation of massive overcontact systems}
\label{sect:contact}

Very close binaries may evolve into contact where both binary
components fill and even overfill their Roche volumes.  The evolution
during the over-contact phase differs from a classical common-envelope
phase as co-rotation can, in principle, be maintained as long as
material does not overflow the L2 point. This means that a spiral-in
due to viscous drag can be avoided, resulting in a stable system
evolving on a nuclear timescale.

As a simple approximation to the modelling of over-contact systems with
a 1D code, we consider the photosphere of such an object to lie on a
Roche equipotential $\Phi$ and divide it into two distinct volumes for each
star, $V_1(\Phi)$ and $V_2(\Phi)$, separated by a plane crossing
through $L_1$ perpendicular to the line joining both stars. We then
associate a volume-equivalent radius to each of these, $R_1(\Phi)$ and
$R_2(\Phi)$, with the radii corresponding to the potential at $L_1$
as the Roche-lobe radius $R_{\mathrm{RL}}$ of each.  When both
stars overfill their Roche-lobes, the amount of overflow of one
component is a function of the mass-ratio, and the amount of overflow
of the other component, i.e.
\begin{eqnarray}
\frac{R_2(\Phi)-R_{\mathrm{RL},2}}{R_\mathrm{RL,2}}=
F\left(q,\frac{R_1(\Phi)-R_{\mathrm{RL},1}}{R_\mathrm{RL,1}}\right),
\end{eqnarray}
where the function $F(q,x)$ must satisfy the conditions $F(q,0)=0$ and
$F(1,x)=x$. By numerically integrating $V_1(\Phi)$ and $V_2(\Phi)$ for
different $q$ ratios and potential values between those at $L_1$ and
$L_2$, we find the fit $F(q,x)=q^{-0.52}x$, with the Roche-lobe radius
computed as in \cite{Eggleton83}, that approximates $F(q,x)$ with a few percent error
in the range $0.1\le q\le 1$. During an over-contact phase,
mass transfer is then adjusted in such a way that the amount of
overflow from each component satisfies this relationship.

Once both stars overflow past the outer Lagrangian point, we expect
the system to merge rapidly, either due to mass loss from $L_2$
carrying a high specific angular momentum or due to a spiral-in due
to the loss of co-rotation. We find that, given the volume equivalent radii
for the potential at the L2 point $R_{\mathrm{L2},i}$, the amount of overflow of the
least massive star in a system that reaches $L_2$ approximately satisfies
the relationship
\begin{eqnarray}
\frac{R_{\mathrm{L2},2}-R_{\mathrm{RL},2}}{R_\mathrm{RL,2}}=0.299\;\tan^{-1}(1.84q^{0.397})
\end{eqnarray}
with an error smaller than $2\%$ in the range $0.02\le q\le1$. At
$q=1$ this means that a star needs to expand up to $1.32$ times its
Roche-lobe radius before reaching L2, leaving a significant amount of
space for a binary to survive through an over-contact phase.  Note
that, for the moment, we ignore the effects of energy and element
transfer through the shared envelope. However, since the systems we
model in this work undergo contact as rather unevolved stars with mass
ratios not far from one, we expect these effects to be of minor
importance for our present study.

It is worth mentioning that VFTS352 is a massive ($\sim
30\mso+30\mso$), short-period ($P_{\rm orb}=1.12$\,d) over-contact
binary, evolving on the nuclear timescale, which has a mass-ratio of
$q=1.008$ and is thought to undergo chemically homogeneous evolution
\citep{Almeida15}.  This system therefore provides direct support for
the MOB scenario and for our treatment of this phase.

\section{Results}
\label{sect:results}

Before looking at the wider parameter space, we provide an example
of a typical MOB evolution in the following section.

\subsection{Exemplary MOB evolution}
\label{sect:examplestar}

\begin{figure}
        \begin{center}
                \includegraphics[width=1.0\hsize]{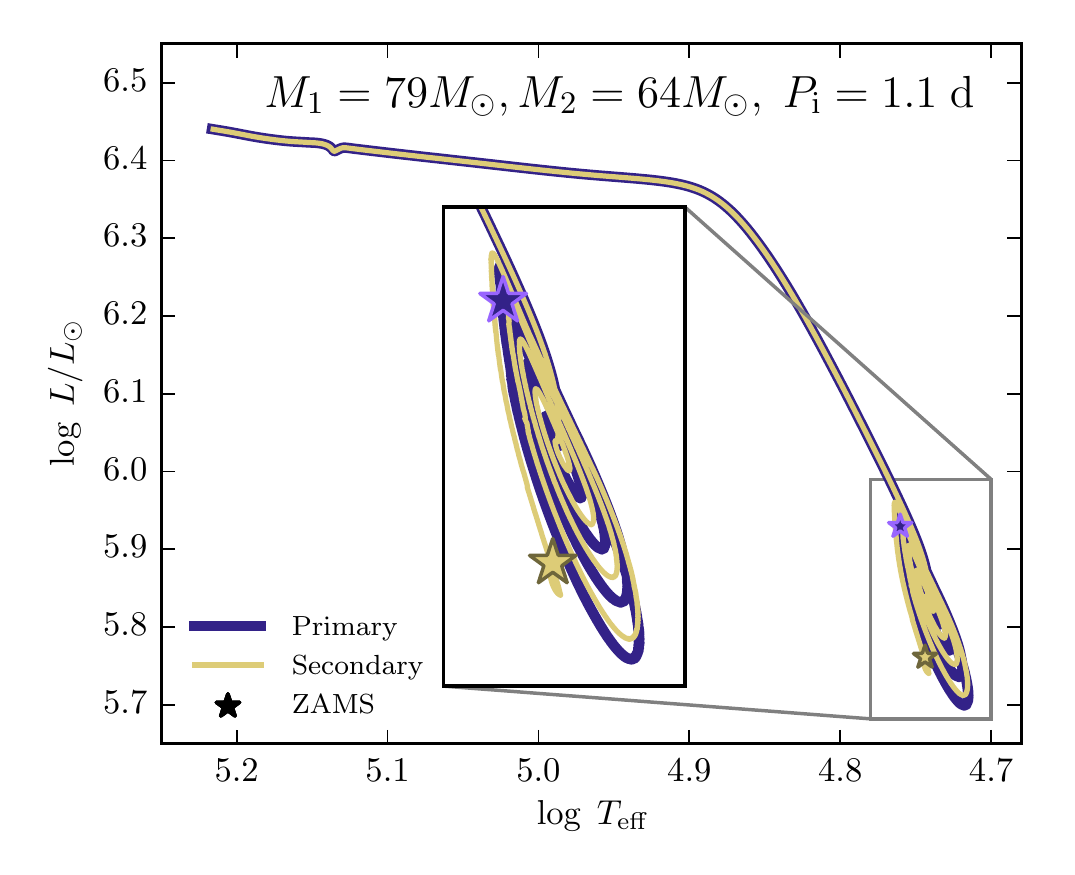}
        \end{center}
        \caption{Evolutionary tracks of both stars in
                a MOB in the HR diagram. The initial masses are
                $79\mso$ and $64\mso$ and the initial orbital period is 1.1\,d.
                Both stars evolve towards nearly equal masses, such that their evolutionary tracks
                after the overcontact phase become almost identical.
        }\label{fig:spiral}
\end{figure}

We show in Fig.\ref{fig:spiral} the evolutionary tracks of both
components of a $79+64\mso$ binary at $Z=\zso/50$ with an initial
period of 1.1\,d, from the zero-age main sequence until core helium
exhaustion. This system enters an overcontact phase early during core
hydrogen burning, during which it swaps mass back and forth several
times. Each time, the mass-ratio becomes closer to one, such that
eventually contact is avoided and the system evolves as a detached
binary with two stars of about $71\mso$ furtheron.

Core hydrogen burning ends at an effective temperature of  
$\log T_{\rm eff}\simeq 4.9$, after which both stars contract to even smaller radii.
Due to the different initial evolution, both stars do not evolve exactly
synchronously during the late evolutionary stages. 

\begin{figure}
        \begin{center}
                \includegraphics[width=1.0\hsize]{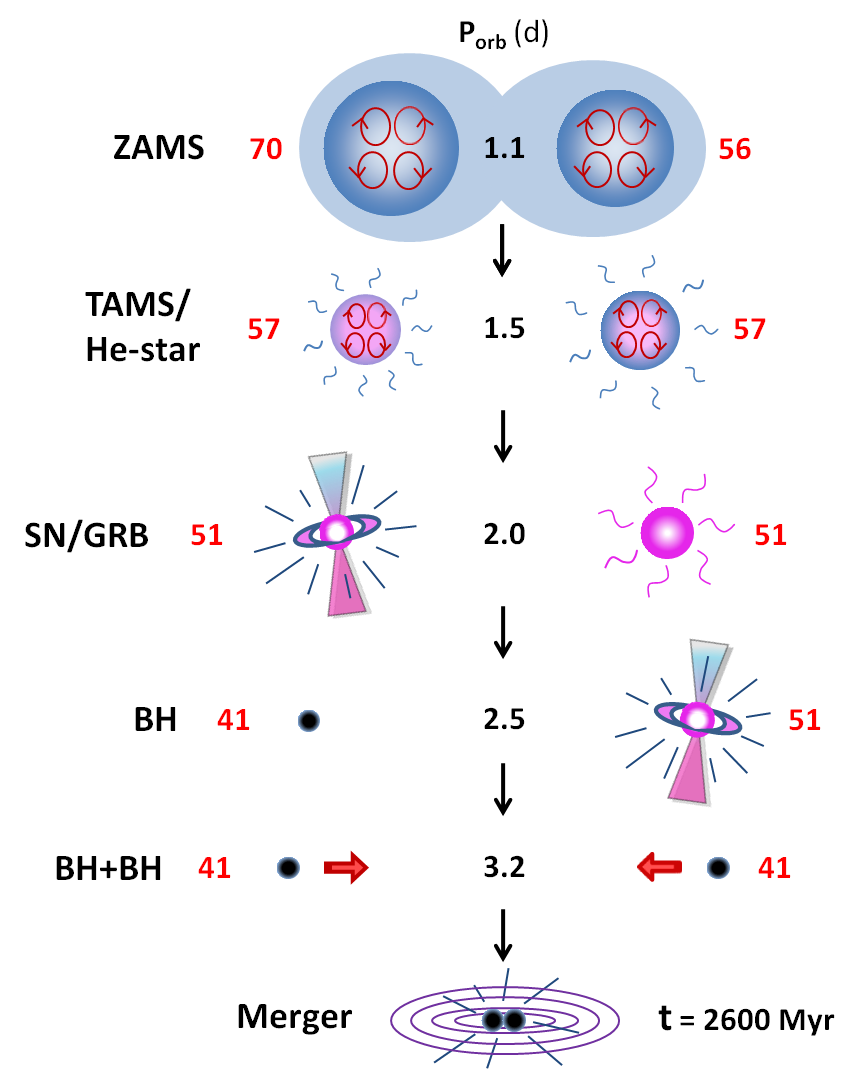}
        \end{center}
        \caption{
                Illustration of the binary
                stellar evolution leading to a BH+BH merger with a high
                chirp mass. {The initial metallicity is $Z_\odot/50$}, the masses of the stars in solar masses are indicated with red
                numbers, and the orbital periods in days are given as black numbers. 
                A phase of contact near the ZAMS causes mass
                exchange.  Acronyms used in the figure. ZAMS: zero-age main
                sequence; TAMS: termination of hydrogen burning; He-star:
                helium star; SN: supernova; GRB: gamma-ray burst; BH: black
                hole.}
        \label{fig:cartoon}
\end{figure}

Figure~\ref{fig:cartoon} illustrates the different evolutionary stages 
of a typical binary in our grid from the zero-age main sequence (ZAMS) to the black-hole
merger stage.  While chemically homogeneous evolution is maintained during all of core hydrogen
burning, again, mass is transferred back and forth
between the two binary components in a succession of contact stages,
which eventually leads to a mass-ratio very close to one.  During the
main-sequence phase and the post-core-hydrogen-burning phase, where
both stars are compact detached helium stars, stellar-wind mass loss
leads to a widening of the orbit. As the winds are metallicity
dependent \citep{Mokiem07,Vink05}, metal-rich systems are found to
widen strongly, which limits the tidal interaction such that the homogeneous evolution
may end. Furthermore, if the orbital period is too long, any BH+BH
binary that may be produced will not merge in a Hubble time.
The depicted case corresponds to case with a metallicity of $Z=\zso/20$,
which provides a black hole merger after 2.6\,Gyr.

\subsection{Example grid}
\label{sect:examplegrid}

\begin{figure}
        \begin{center}
                \includegraphics[width=1.0\hsize]{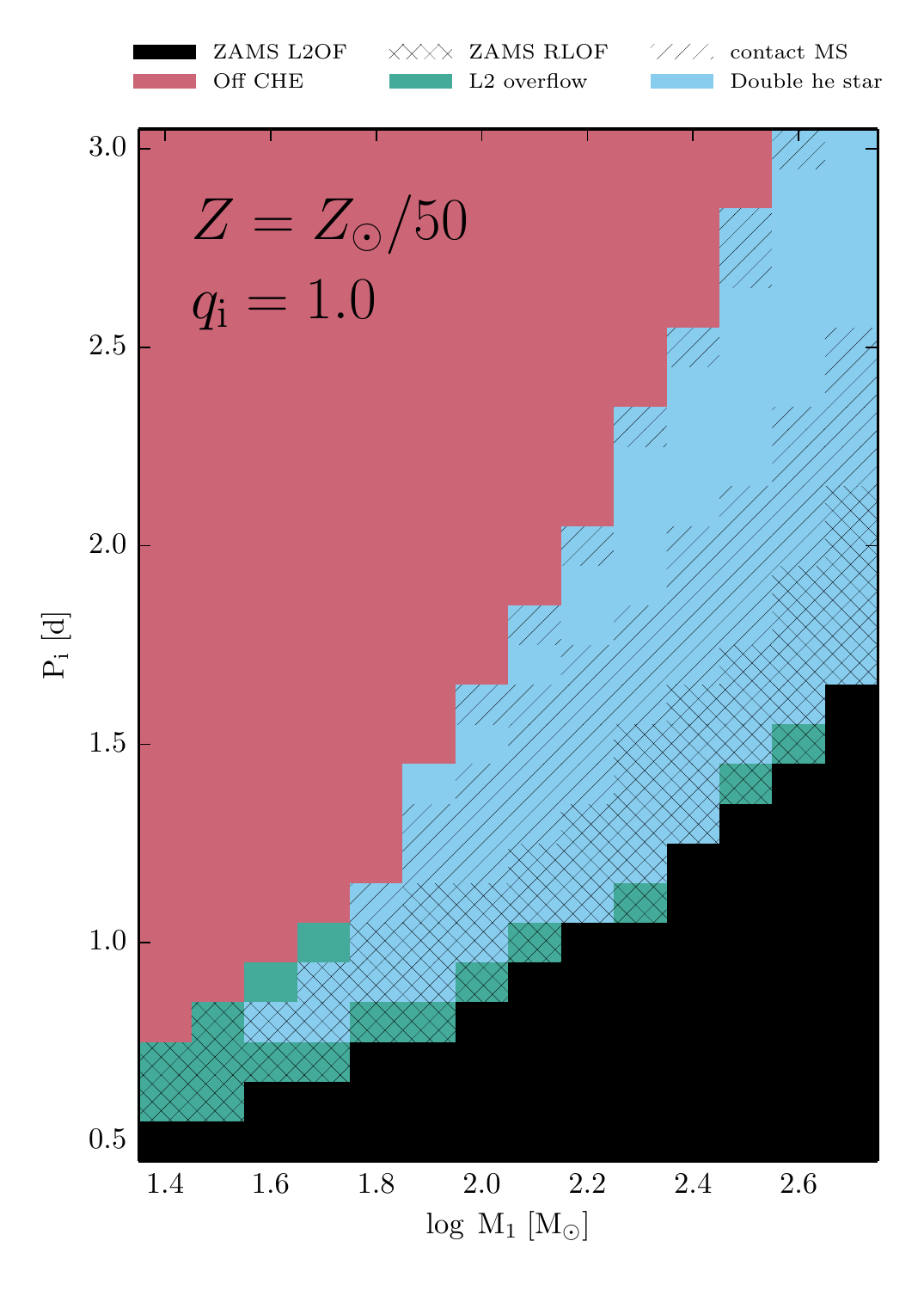}
        \end{center}
        \caption{ Example of a grid of binary systems (initial orbital
                period versus initial primary mass) with $Z=Z_\odot/50$ and
                $q_\mathrm{i}=1$. Models that reached a point where one of the stars
                has a difference between the surface and central helium abundance
                of more than $0.2$ are considered not to be evolving
                chemically homogeneously and their calculation is stopped
                (pink color). The region where the initial orbital period is small
                enough as to have L2 overflow at the ZAMS is marked in black, while
                those systems that reach L2 overflow during the main sequence are marked in green.
                Systems marked in blue successfully form double helium stars. Single
                hatching marks systems which experience contact during the main sequence, while
                doubly hatched ones are in an overcontact phase already at the ZAMS.
        }\label{fig:grid}
\end{figure}

An example of a grid of binary systems is shown in 
Fig.\,\ref{fig:grid}, for $Z=Z_\odot/50$ and $q_\mathrm{i}=1$. 
Each rectangle in the plot corresponds to one detailed binary evolution model.

As Fig.\,\ref{fig:grid} shows,
progenitors of massive double helium stars require 
initial primary masses above $\sim
30\mso$, and the range of periods for which they are formed widens
with increasing primary mass.  This broadening 
is the consequence of the larger convective cores and stronger
winds for the more massive stars; this has a similar effect as the
rotational mixing in exposing helium-rich material at the surface
\citep{Kohler15,Szecsi15}.

In particular for binaries with final masses low enough to
avoid the pair-instability regime (i.e., roughly for $M_{\rm f} < 60\mso$;
cf. Sects.\,\ref{sect:collapse} and \ref{sect:kick}),
the parameter space for progenitors is completely
dominated by overcontact systems, with an important
fraction coming from systems that have such low periods that they
already overflow their Roche-radii at the ZAMS. 
Only for initial masses well above $100\mso$ do we find systems
which avoid contact throughout their evolution. But, even in 
this mass regime, most systems go through at least one contact phase.  

We stop all but three (see Sect.\,\ref{sect:collapse})
of our binary-evolution models at a time when
the stars end core helium burning since their fate is settled at that
time, and the binary orbit will essentially not change any more until
the first stellar collapse occurs (3rd stage in
Figure~\ref{fig:cartoon}).

\subsection{Final binary configurations}
\label{sect:periods}

\begin{figure}
        \begin{center}
                \includegraphics[width=1.0\hsize]{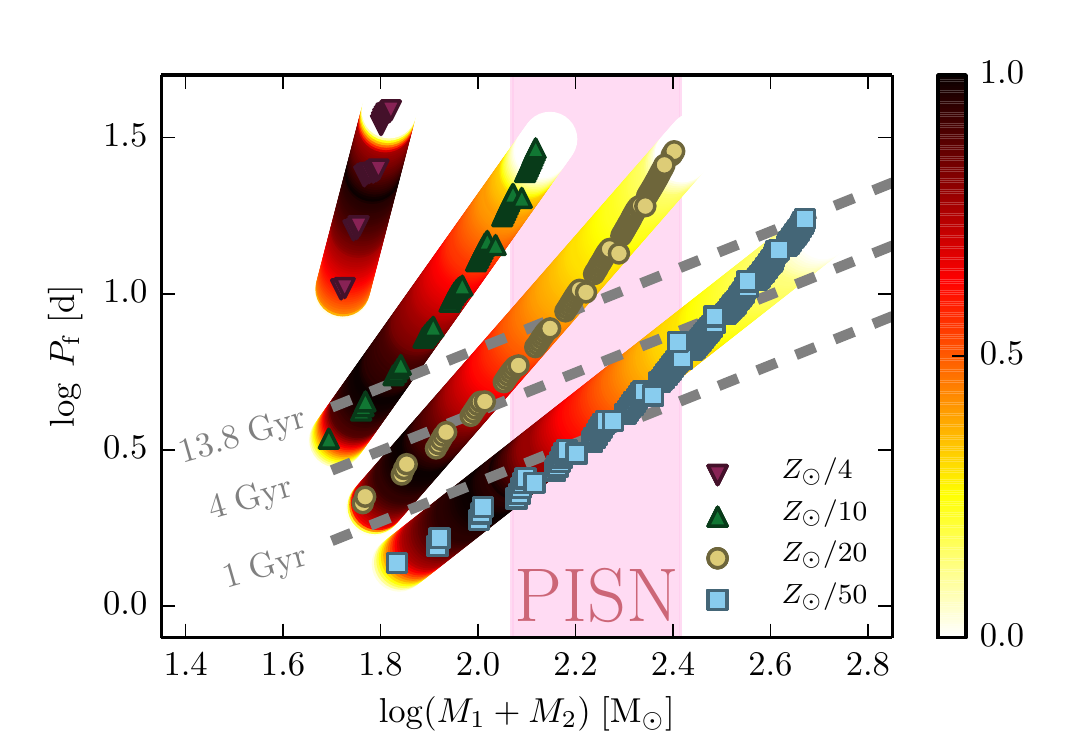}
        \end{center}
        \caption{
Total masses and {orbital} periods at core helium depletion for systems with $q_\mathrm{i}=1$ at four
different metallicities. Dashed lines are for constant merger times assuming
direct collapse into a black hole, and the shaded region indicates the mass
range at which PISNe would occur, resulting in the total disruption of the stars
instead of black-hole formation. The colored bands represent for each
metallicity the relative number of objects formed.
                \label{fig:MP}
        }
\end{figure}

Figure~\ref{fig:MP} summarizes the
distribution of the final total system masses as a function of
their final orbital period for those models in our grid 
which succeeded in producing
close pairs of helium stars. Since the initial binary periods have
to be very short in order to enforce the rapid rotation required for
homogeneous evolution, the final properties lie in a narrow strip for
each metallicity, but these are distinctly different for different
metallicities. For the highest considered masses, 
this is mainly due to the metallicity dependence of the
stellar wind mass loss which has the effect of widening the systems and reducing
the mass of the stars, thus producing systems with longer orbital
periods and lower masses at higher metallicity. 

Figure~\ref{fig:MP}
also indicates the merger times for these systems, assuming that the
masses and periods do not change in the black-hole formation process
(cf., Sect.\,\ref{sect:kick}). All models with $\zso /4$ and all but
the lowest-mass ones with $\zso /10$ produce binaries which are too
wide to lead to black-hole mergers within a Hubble time.  
The more metal-poor models, on
the other hand, produce very tight He-star binaries below as well as
above the mass regime where pair-instability supernovae (PISNe) are
expected to lead to the complete disruption of the stars rather than
the formation of black holes \citep{Heger02,Chatzopoulos12}.

The trend of shorter merger times for lower metallicities is expected to continue
towards the lowest metallicities found in the Universe. As stellar wind mass loss
becomes more and more negligible, the initial stellar radii determine the smallest
possible orbital periods. As an example, stars of $60\mso$ have ZAMS radii of
$12\rso$, $10.5\rso$, $10\rso$, and $3.5\rso$, at $Z=\zso$, $\zso/10$, $\zso/50$
and $Z=0$, respectively. This implies that the merger times for the lowest metallicities,
in particular for Population\,III stars, become extremely short. While the expected number
of such objects is small, this opens the exciting possibility of eventually observing primordial
black hole mergers at high redshift.

\subsection{Mass distribution and mass-ratios}
\label{sect:masses}

\begin{figure}
        \begin{center}
                \includegraphics[width=1.0\hsize]{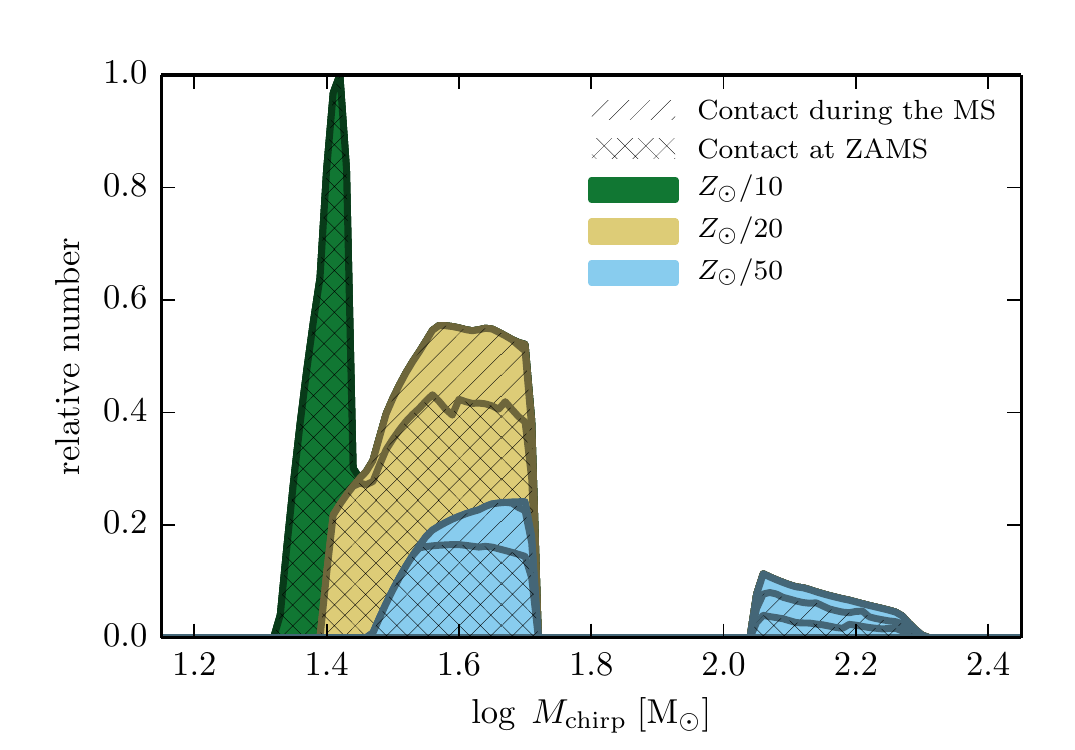}
        \end{center}
        \caption{{Stacked} distribution of chirp masses of BH+BH
                systems formed at different metallicities, such that
                they merge in less than $13.8\,\mathrm{Gyr}$. The
                contribution from each metallicity is scaled
                assuming a flat distribution in $Z$. At very short
                periods, systems are already at contact at the ZAMS.
        }\label{fig:chirp}
\end{figure}

Figure~\ref{fig:chirp} shows the predicted intrinsic chirp-mass
distribution for BH+BH mergers for our different metallicity grids,
again assuming no mass loss in the BH-formation process.  The most
prominent feature is the prediction of a clear gap in this
distribution, which occurs because systems which would otherwise
populate this gap do not appear since the stars explode as
pair-instability supernovae without leaving a stellar remnant. The BH
progenitors in the systems above the gap also become pair unstable,
but the explosive burning can not reverse the collapse which leads
straight to the formation of a black hole \citep{Heger02,Langer12}.

There is a
strong general trend towards higher chirp masses with decreasing
metallicity. At the lowest metallicity ($Z=Z_\odot/50$) we 
produce also BHs above the PISN gap. While obviously their number is
smaller than the number of BH systems below the gap, they may still be
significant as the amplitude of the gravitational-wave signal
is a strong function of the chirp mass (cf. Sect.\,\ref{sect:rates}).

\begin{figure}
        \begin{center}
                \includegraphics[width=1.0\hsize]{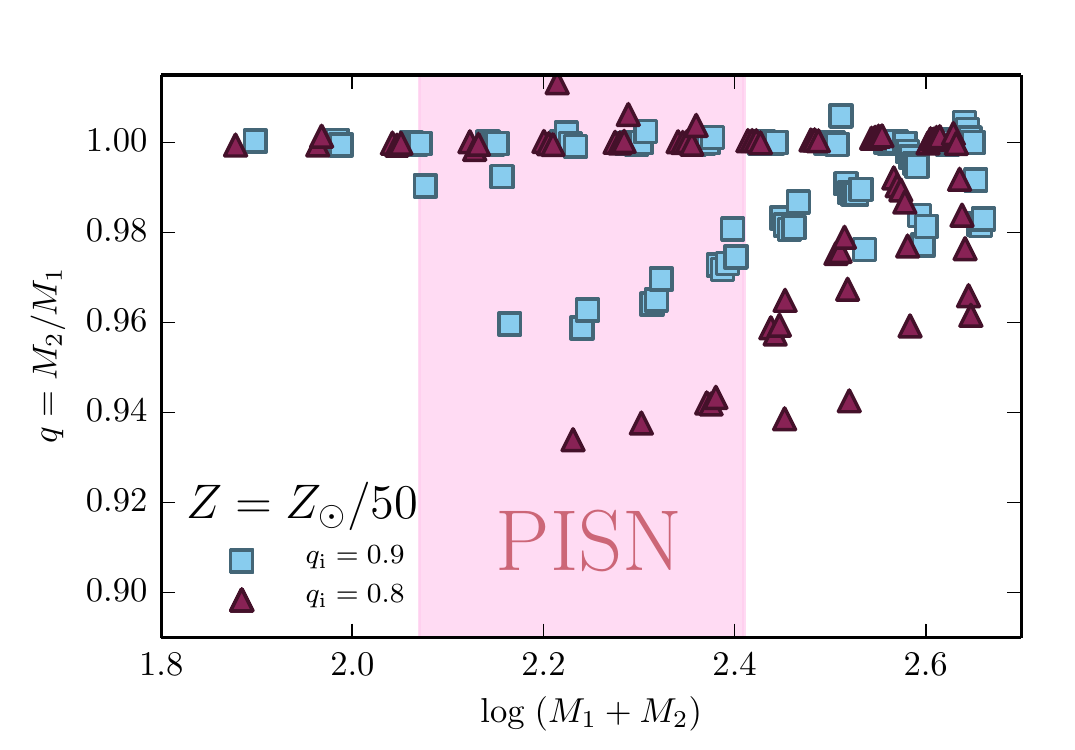}
        \end{center}
        \caption{Mass-ratios of BH+BH systems resulting from our modelled systems for
                $q_\mathrm{i}=0.9$ and $q_\mathrm{i}=0.8$ and a metallicity $Z=Z_\odot/50$ under the
                assumption that no mass is lost during collapse. The shaded region
                indicates the limits for the occurrence of PISNe.
                }
        \label{fig:mratios}
\end{figure}

As indicated in Fig.\,\ref{fig:chirp}, the vast
majority of merging systems have passed through a contact phase. Since both stars are relatively
unevolved when they undergo contact, these contact phases result in
mass transfer back and forth until a mass-ratio $q\simeq 1$ is
achieved. This is depicted in Figure \ref{fig:mratios}, where final
mass-ratios are shown for systems with $q_\mathrm{i}=0.9,0.8$ and $Z=Z_\odot/50$. For each mass-ratio,
two distinct branches are visible, corresponding to systems that
undergo contact and evolve to $q\simeq 1$, and systems that avoid
contact altogether. Owing to the strong dependence of mass-loss rates
with mass, at high masses, even systems that avoid contact altogether
evolve towards $q=1$. 

{\cite{MandelMink16} model this channel without including contact systems 
and find an important number of binaries forming double BHs from progenitors 
below the PISN gap, with final mass-ratios in the range of 0.6 to 1,
reflecting just a small shift from the initial mass-ratio distribution
due to mass loss. 
However, \cite{MandelMink16} do not perform detailed stellar evolution calculations.
They check whether their binary components underfill their Roche-radii at the ZAMS,
and then assume that this will remain so in the course of the quasi-homogeneous evolution
of both stars. When considered in detail, however, in particular the more massive and
more metal-rich stars undergo some expansion during core hydrogen burning, even on the
quasi-homogeneous path \citep{Brott11,Kohler15,Szecsi15}, likely due to the increase
of their luminosity-to-mass-ratio and the related approach to the Eddington limit
\citep{Sanyal2015}. As a result, the vast majority of the binaries
considered by \cite{MandelMink16} enter contact when computed in detail. Therefore,
our final mass-ratio distribution is much more strongly biased towards $q=1$.
} 

\subsection{Merger Delay Times}

\begin{figure}
	\begin{center}
		\includegraphics[width=1.0\hsize]{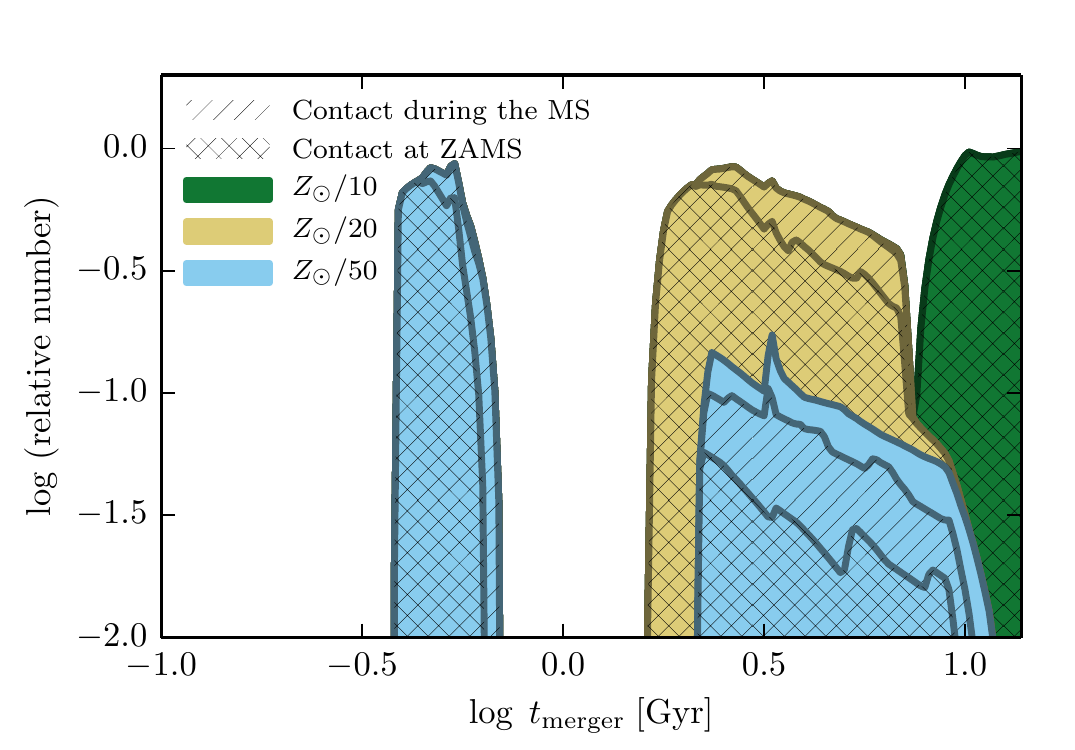}
	\end{center}
\caption{{Stacked} distribution of merger delay times for different metallicities
	(as indicated). The meaning of the hatching is as in Figure \ref{fig:chirp}}\label{fig:delay}
\end{figure}

As Fig.~\ref{fig:MP} already indicates, the merger delay time,
i.e. the time between the formation of the BH+BH binary and the
eventual merger, is a strong function of metallicity, where the merger
delay times (at a given BH mass) are systematically shorter for lower
metallicity. Fig.~\ref{fig:delay} shows the distribution of the merger
delay times for the different metallicities in our grids (assuming a
uniform metallicity distribution). While the typical delay time is
several Gyr, which helps detecting these events at lower redshift (see
Sect.~\ref{sect:rates}), at the lowest metallicity, the delay time can
be as short as $0.4$\,Gyr for BH+BH mergers below the PISN
gap. 

{The decrease in delay times with smaller metallicity is not found in the
models of \citet{MandelMink16}, who conclude that  no high-redshift mergers are expected. 
The reason is that they consider effectively only
one metallicity, namely the threshold metallicity for chemically homogeneous
evolution by \citet{Yoon06} of $Z=0.004$. However, the components of
lower-metallicity binaries are more compact, allowing tighter binaries
at zero age, and have weaker winds, which produces tighter double-black-hole binaries.
Therefore, while \citet{MandelMink16} predict delay times to be larger than
3.5\,Gyr, we find up to 10-times smaller values at our lowest metallicity
(Figure \ref{fig:delay}). Since the minimum delay time depends on the metallicity-dependent
stellar radii, and stellar radii of massive metal-free stars are less than half
compared to those at $\zso/50$ \citep{Yoon2012,Szecsi15},
even orders of magnitude shorter merger times can be expected. Therefore, even though 
much rarer, we argue that massive BH mergers could occur up to the redshift of
Population~III stars.} If such mergers were detected, it would allow us to probe
the evolution of massive stars in the very early Universe.

We also note that, if the black holes receive kicks at birth,
even higher-metallicity systems may merge very rapidly if the kick
reduces the pericenter distance (see Appendix ~\ref{sec:BHkick}).

\subsection{Observational counterparts}
\label{sect:obs}

Our choice to include models of up to 500$\mso$ in our grids is
supported by the evidence for stars of such high masses in the LMC
\citep{Crowther10}.  

Various evolutionary stages of the MOB scenario are
observationally confirmed by massive binary systems in nearby
galaxies. As mentioned in Sect.\,2, the MOB VFTS352 
\citep{Almeida15} supports the idea of
homogeneous evolution of overcontact binaries, even though it is not expected
to lead to a black-hole merger due to the rather high metallicity
of the LMC (cf., Fig.\,\ref{fig:MP}). It corresponds to
the first stage of our cartoon in Fig.\,\ref{fig:cartoon}.

The SMC binary HD\,5980 corresponds well to the second stage of
Fig.\,\ref{fig:cartoon}. It consists of two $\sim 60\,\mso$ stars
which are both very hydrogen-poor, in a $\sim$19\,d orbit.
\citet{Koenigsberger14} concluded that this system most likely emerged
from homogeneous evolution. This system is well recovered in our grid
at $\zso/10$.

Finally, IC10\,X-1
and NGC\,300\,X-1 are binaries which may correspond well to stage~3
of Fig.\,\ref{fig:cartoon}. Both have a short orbital period
($P_{\rm orb}\simeq 1.5$\,d for both) and contain very massive
black-hole primaries ($> 23\mso$ and $20\mso$) and similar-mass,
hydrogen-free companions ($\sim 35\mso$ and $26\mso$;
\citet{Barnard08,Bulik11}).  Both systems have close-matching
counterparts in our $\zso/20$ binary-evolution grids, 
{with life times of up to several 10$^4$\,yr.}

\subsection{Spins}
\label{sect:spins}

\begin{figure}
        \begin{center}
                \includegraphics[width=1.0\hsize]{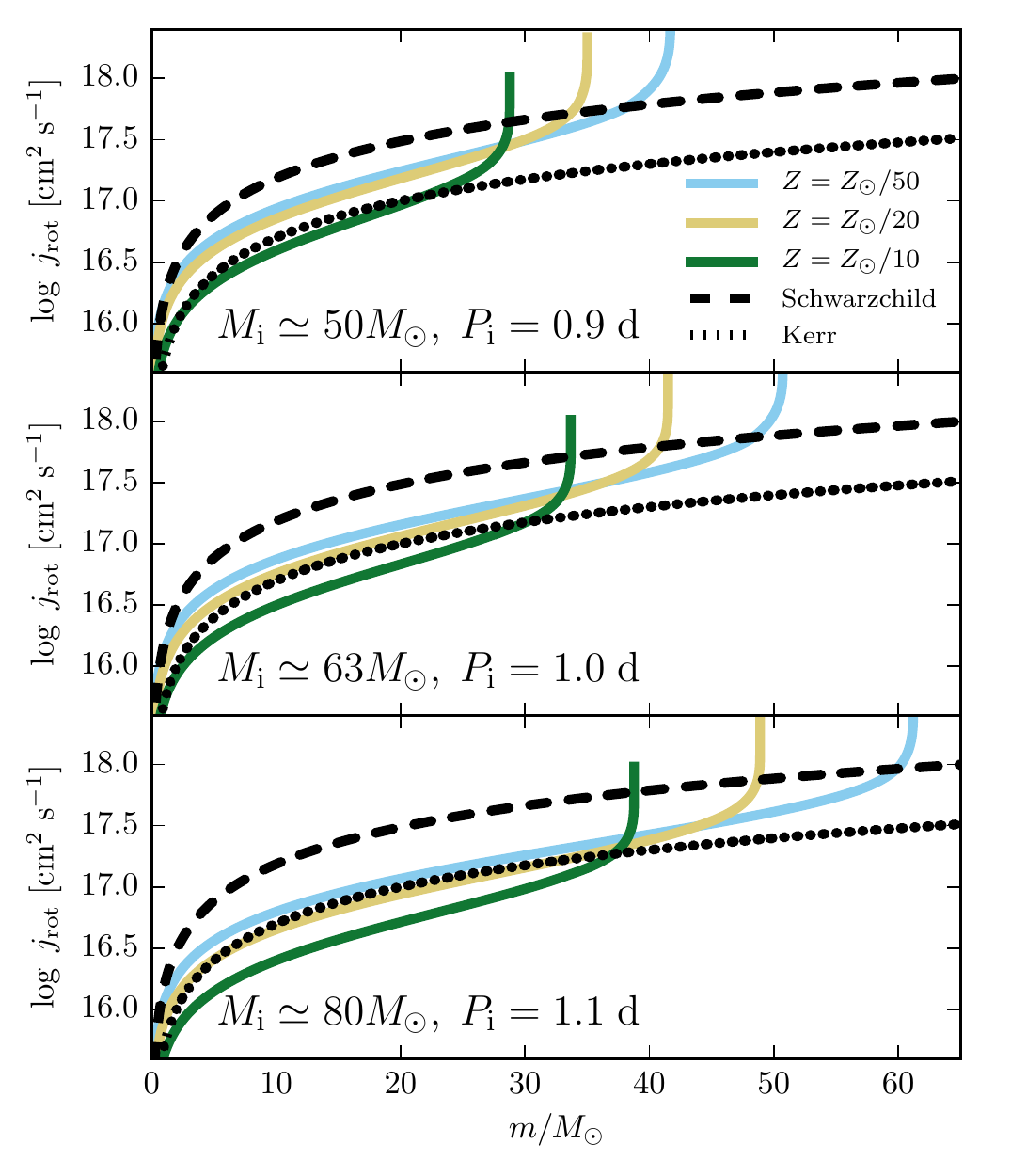}
        \end{center}
        \caption{Angular-momentum profiles at core helium depletion for
                the primary stars of binaries from our grid 
                that result in double-helium-star binaries. Shown are stars
                of three different initial masses 
                in binaries with similar initial orbital periods, at metallicities of
                $Z=Z_\odot/50,Z_\odot/20,Z_\odot/10$. The curves
                for the specific angular momentum of the last stable orbit for
                a non-rotating (Schwarzschild) and critically rotating (Kerr) black
                hole are also included.
        }\label{fig:am}
\end{figure}

To test the possibility of producing LGRBs according to the
``collapsar'' scenario \citep{Woosley93} from our MOB models, we
compare in Fig.\,\ref{fig:am} angular-momentum profiles at the point
of core helium depletion for a few systems that fall below the PISN
gap. A significant amount of mass ejected during an LGRB event could
modify the final orbital periods of double BHs, although we find that
this should not play a determining role in our rate estimates (as
discussed further in Sect.~4).

\begin{figure}
        \begin{center}
                \includegraphics[width=1.0\hsize]{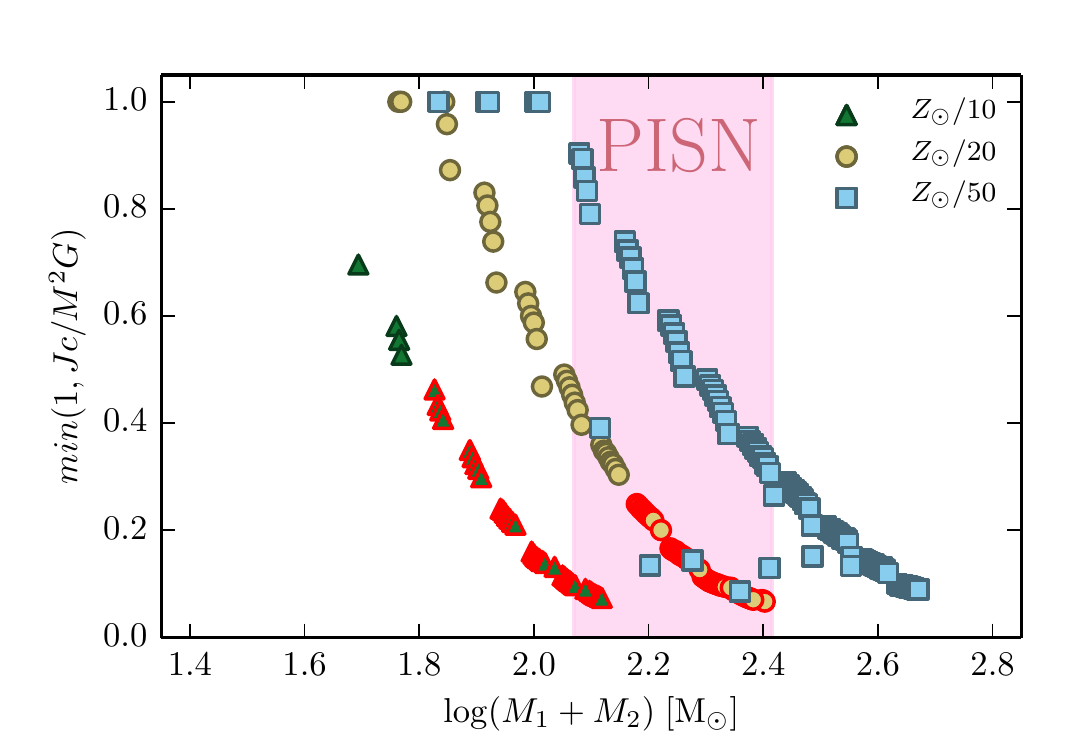}
        \end{center}
        \caption{Kerr parameter as function of the final system mass,
                for our models at $Z=Z_\odot/50,Z_\odot/20,Z_\odot/10$,
                assuming a complete collapse of our helium stars to black holes.
                Binaries indicated through symbols with a red frame have
                merger times which exceed the Hubble time.
        }\label{fig:kerr}
\end{figure}

As Fig.\,\ref{fig:am} shows, 
models at a metallicity of $Z_\odot/10$ experience
significant braking due to winds, and thus they are unlikely to
produce LGRBs. In contrast, several systems at $Z_\odot/50$ that
result in helium stars below the PISN gap have specific
angular-momentum profiles above the values for the last stable orbit,
assuming all mass collapses into a critically rotating black hole.
The results at $Z_\odot/20$ are more ambiguous, and it is not clear
whether the stars would produce an LGRB or not. For systems forming
black holes above the PISN gap, even for low metallicity wind braking
is strong enough to avoid the formation of LGRBs.
This picture is confirmed when considering the Kerr parameters of our models
in Fig.\,\ref{fig:kerr} in the different mass and metallicity regimes.

\subsection{Models up to core collapse}
\label{sect:collapse}

\begin{figure}
        \begin{center}
                \includegraphics[width=1.0\hsize]{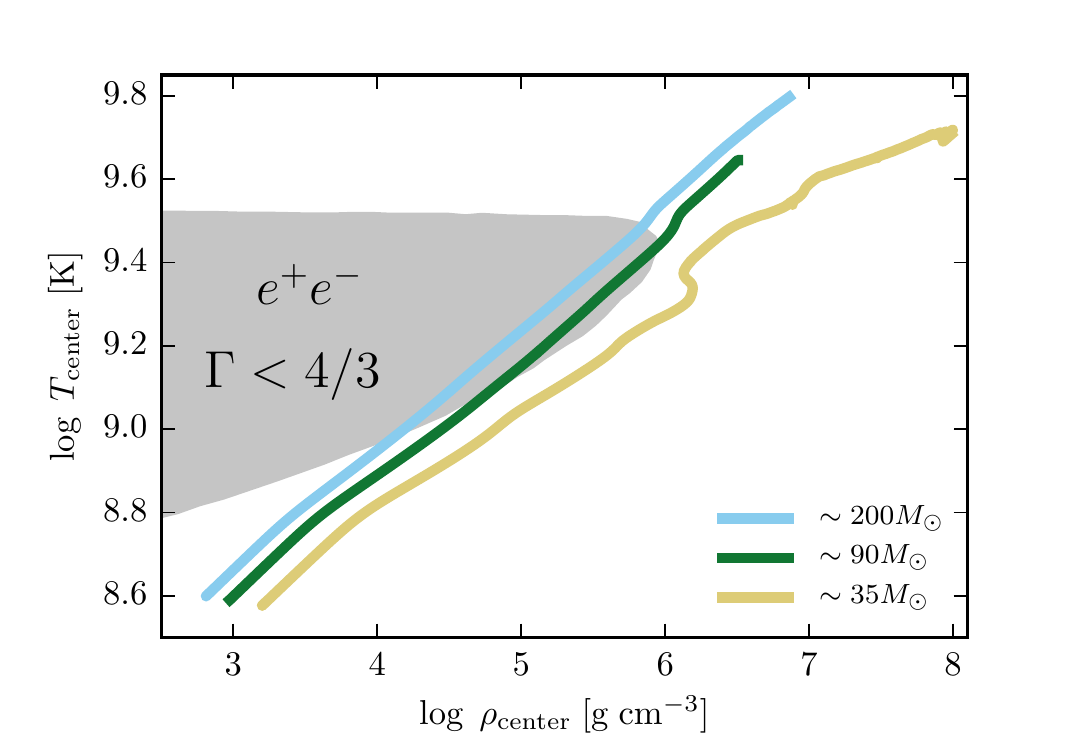}
        \end{center}
        \caption{ The evolution in the $T_c - \rho_c$-diagram for the
                three stellar models at $Z=Z_\odot/50$ (with the masses at helium depletion as indicated)
                calculated to the final evolutionary stage. The shaded
                region shows the region that is unstable to pair
                creation. Both the 35\,$M_\odot$ and the 200\,$M_\odot$ stars collapse
                to form black holes, while the 90\,$M_\odot$ is disrupted in
                a PISN.
        }\label{fig:trho}
\end{figure}

To depict the effect of the PISN gap we took three models with masses
$200\,M_\odot$, $90\,M_\odot$ and $35\,M_\odot$ and metallicity
$Z_\odot/50$ after helium depletion and evolved these through the late
evolutionary phases. Figure \ref{fig:trho} shows the evolution of
central density and temperature of each star, together with the region
at which pair production results in an adiabatic index of $\Gamma < 4/3$. 

The least massive of the three
stars avoids the pair-unstable region altogether and experiences core
collapse after silicon depletion. In the $90\,M_\odot$ model, the core
collapses during oxygen burning, resulting in explosive burning which
injects enough energy to halt the collapse and drive an explosion.  At the
highest mass, the oxygen core also undergoes collapse, but explosive
burning is not sufficient to stop it, and in the end burning proceeds
very fast up to silicon depletion, resulting in an iron core with an
infall velocity $>1000\,\mathrm{km\;s^{-1}}$.

\subsection{Explosive mass loss and momentum kicks}
\label{sect:kick}

In all models below the pair-instability regime we expect the
formation of black holes. If the whole star collapses without ejecting
any mass or energy, the masses and periods in Figure~\ref{fig:MP}
would also represent the masses of the final black holes and the
post-collapse orbital periods. On the other hand, as our helium stars
tend to be rapidly rotating, some of them may go through a collapsar
phase \citep{Woosley93}, producing LGRBs, in which part of the collapsing star is ejected, and the
binary orbit may receive a supernova kick. The effect of the mass loss
would be to reduce the final black-hole masses (and to reduce the
strength of any eventual gravitational-wave signal) and widen the
system (and increase the merger time), while the effect of a kick can
be to either increase or decrease the orbital period and the merger
time (see Appendix\,\ref{sec:BHkick} for a more detailed
discussion). While the details of the collapse phase are still very
uncertain, which may have an effect on the BH+BH detection
rates, our main conclusions are not dependent on these. 

In any case,
the final angular-momentum profiles of our models
(see Sect.\,\ref{sect:spins}) suggest that only the lowest-mass models
($M_{\rm final} \simle 40\mso$) at the two lowest metallicities ($Z=\zso
/20$, $\zso /50$) may retain enough angular momentum in the core to be
good LGRB candidates. Nevertheless, because of the large amount of
available angular momentum, we expect many of the BHs formed in this
scenario to be rapidly rotating, with the spin parameter
roughly scaling inversely with the final orbital period shown in
Fig.~\ref{fig:MP} (i.e.\ the largest spins are expected for the
lowest-mass BHs at the lowest metallicity).  Finally, we note that,
below the disruptive PISN regime, there is a regime of pulsational
PISNe \citep{Chatzopoulos12}, where substantial mass loss is expected
but a BH is nevertheless ultimately formed \citep{Woosley07}.

\section{Merger rates}\label{sect:rates}


{Concerning the conventional scenario to produce close double compact binaries involving common envelope evolution (see Appendix\,B),} except for a few cases \citep{VossTauris03,Belczynski+10,Dominik+15}, the far majority of published population synthesis studies predict a much higher NS+NS merger rate per Milky~Way equivalent galaxy (MWEG) compared to the rate of BH+BH mergers. Based on a detailed comparison study of published models \citep{Abadie10}, the NS+NS merger rate was estimated to be $100\,{\rm MWEG}^{-1}\,{\rm Myr}^{-1}$, which is about 100 times larger than the rate predicted for BH+BH binaries. However, given the more massive compact objects in BH+BH binaries, compared to those of NS+NS binaries, their emitted gravitational-wave amplitudes are significantly larger {such that the LIGO detection rates of both come out to be approximately equal. The so-called 'realistic' rates	quoted by \citet{Abadie10} are 40 and 20$\,{\rm yr}^{-1}$ for NS and BH mergers, respectively, but the uncertainty in these numbers is larger than three orders of magnitude.}

\subsection{Chirp masses and MOB evolution}
Two important things should be kept in mind with regard to these
quoted rates.  First, the BH+BH binaries are assumed to be composed of
$10\,{\rm M_{\odot}}$~BHs (even $5\,{\rm M_{\odot}}$~BHs in all LIGO result papers
published before 2010), corresponding to an intrinsic chirp mass,
$\mathcal{M}_0$ of $\sim\!8.7\,{\rm M_{\odot}}$ (for equal mass binaries,
$\mathcal{M}_0=(1/4)^{3/5}\,M\simeq 0.435\,M$, where the total mass,
$M$ is twice the BH mass, $M_{\rm BH}$). For low metallicities, our
MOB scenario predicts formation of BHs with masses of
$25-60\,{\rm M_{\odot}}$ and $130-230\,{\rm M_{\odot}}$ (i.e. below and above the
PISN mass range, respectively), resulting in very large intrinsic
chirp masses of about $20-50\,{\rm M_{\odot}}$ and $115-200\,{\rm M_{\odot}}$,
respectively (cf. Fig.~\ref{fig:chirp}).  Such mergers can be seen
throughout a significant fraction of the Universe, since the distance
luminosity, $d_L \propto \mathcal{M}_0^{5/6}$.  Note, the detected
chirp masses will be redshifted to $\mathcal{M}_0\,(1+z)$, where $z$
is the BH+BH system's redshift with respect to the detector on Earth
\citep{Finn96}.

Secondly, all previously published rates were based on CE-evolution (cf., App.\,\ref{sec:BHstd}),
which creates uncertainties in the rates by more than two orders of
magnitude as a result of our poor understanding of the CE physics \citep{Dominik+12}. The new BH+BH formation scenario (MOB evolution) 
presented in this work does not involve any CE phase. Instead, it is based on much less uncertain physics (as discussed in the previous section).
Equally important, it leads to the formation of much more massive BHs compared to previous studies.

Assuming, as a first approximation, that the detection rate,
$\mathcal{R}$, scales with $d_L^{\;3}\propto \mathcal{M}_0^{5/2}$, we
conclude that the expected LIGO detection rates for these massive
BH+BH binaries could easily dominate the overall rates; they are
therefore excellent candidates for the first LIGO source detection
(see the more detailed discussion below).  It should be noted that
some previous studies \citep{Belczynski+10,Dominik+15,Rodriguez15}
have already alluded to a dominance of relatively massive BH+BH
mergers in a low-metallicity environment (or, in particular, via
dynamical channels in dense clusters), although without a specific
detailed model for the binary case.

The expected LIGO detection rate of BH+BH binary mergers has been estimated in the following manner, 
$\mathcal{R}=r_{\rm MW}\times N_{\rm gal}$, where $r_{\rm MW}$ is the expected merger rate in a MWEG, and
\begin{equation}
N_{\rm gal} = \frac{4}{3}\,\pi\,\left(\frac{d_{\rm horizon}}{{\rm Mpc}}\right)^3 
\,(2.26)^{-3}\,(0.0116)
\label{eq:Ngal}
\end{equation}
is the number of MWEGs out to a horizon distance, $d_{\rm horizon}$
\citep{Abadie10}. Here the factor 1/2.26 is included to average over
all binary orientations and sky locations, i.e.  $d_{\rm
  horizon}=2.26\,d_{\rm avg}$ \citep{FinnChernoff93}, and $1.16\times
10 ^{-2}\,{\rm Mpc}^{-3}$ is the extrapolated space density of MWEGs
\citep{Kopparapu+08}. For relatively low-mass BH+BH mergers, assuming
$M_{\rm BH}=10\,{\rm M_{\odot}}$ and a corresponding average design
distance luminosity of $d_{\rm L}\simeq1000\,{\rm Mpc}$ for advanced
LIGO (aLIGO), the estimated values are \citep{Abadie10}: $r_{\rm MW}=0.4\,{\rm
  Myr}^{-1}\,{\rm MWEG}^{-1}$ and $\mathcal{R}=20\,{\rm yr}^{-1}$ .
For a massive BH+BH merger with $M_{\rm BH}=60\,\mso$ (or
$130\,\mso$), we get $d_L\simeq 4.5\,{\rm Gpc}$ (or $d_L\simeq
8.5\,{\rm Gpc}$), and thus $d_{\rm horizon}\simeq 10\,{\rm Gpc}$ (or
$d_{\rm horizon}\simeq 19\,{\rm Gpc}$).  The expected redshift is
$z=1.4$ (or $z=2.3$) for standard cosmological parameters
($H_0=69.6\,{\rm km\,s}^{-1}\,{\rm Mpc}^{-1}$, $\Omega_{\rm M}=0.286$,
$\Omega_{\rm vac}=0.714$).

\subsection{Model assumptions for estimating aLIGO detection rates}

To calculate the aLIGO detection rate of such massive BH+BH
mergers, we first need to calculate the intrinsic merger rate in a
MWEG.  For a given metallicity, it is straightforward to calculate the
rate for various events from our binary evolution grids. We use
simple, standard assumptions about the initial binary parameters: we
assume that (1) the orbital period distribution is flat in $\log P$
and covers the range ($0.5\,\mathrm{d}-1\,\mathrm{yr}$), (2) the
primary mass distribution is described by a Salpeter power law
($dN/d\log M\propto M_1^{-1.35}$), (3) the mass-ratio distribution is
flat, (4) stars more massive than $8\,M_\odot$ produce a core-collapse
SN, and (5) that there is one binary system for every 3~core-collapse
SNe {(i.e. two out of three massive stars are formed in close binaries)}.

Guided by the results from our grids, we require that the initial mass-ratio has to be larger than 0.8 to ensure chemically homogeneous
evolution for both stars. With these assumptions we can calculate the fraction of systems that produce BH+BH mergers relative to the
core-collapse SN rate. These are shown in the first two rows in Table~\ref{table:rates} for various metallicities for BH+BH mergers
below and above the PISN gap. These number imply that for, $Z<Z_\odot/10$, there is typically one BH+BH merger event for $\sim 1000$ core-collapse SNe. 
To relate these numbers to a rate for a MWEG, we need to multiply these fractions by the core-collapse rate for a MWEG. 
Adopting a typical rate of 0.01\,yr$^{-1}$, this implies, e.g., that in a MWEG with $Z=Z_\odot/50$, the BH+BH merger rates are
$6.7\,{\rm Myr}^{-1}$ and $2.7\,{\rm Myr}^{-1}$ below and above the PISN gap, respectively. Note that our model only predicts BH+BH
mergers above the PISN gap at the lowest metallicity.

\subsection{Accounting for the star-forming history and the galactic metallicity distribution throughout the Universe}\label{subsubsec:mapping}

The actual aLIGO detection rate depends on two main factors: 
(1) the detection volume within which a particular event can be detected, and
(2) the cosmological distribution of the sources which depends on the star formation history and the chemical evolution of the Universe.
The detection volume can be estimated using Eq.~\ref{eq:Ngal} where $d_{\rm horizon}$ depends on the aLIGO sensitivity and the BH masses. 
However, Eq.~\ref{eq:Ngal} is only a good approximation for $d_{\rm horizon}\simle 1\,$Gpc as it ignores cosmological expansion. 
To take this into account (in a very approximate way), we introduce a simple cut-off of $N_{\rm gal}=10^{10}$ in Eq.~\ref{eq:Ngal}. 
For reference, this implies that our simple model assumes that there are about 3 core-collapse SNe per second in the Universe. 
Scaling the horizon distance of the BH+BH masses produced in our grids to the design sensitivity of aLIGO, we start by calculating the aLIGO 
detection rate assuming that all galaxies have a particular metallicity. These rates are shown in the last two rows of Table~\ref{table:rates},
in columns 2--5.

One of the main problems is that our rates are a strong function of metallicity and therefore depend on the evolution of metallicity with
time and the spread of the metallicity distribution at a given redshift. For example, the mean metallicity of galaxies will only be
less than $Z_\odot/50$ at redshifts larger than $5-7$ (A.~Fruchter; private communication). On the other hand, even in the local Universe
there are some galaxies (mostly dwarf galaxies) with extremely low metallicities. A proper calculation of this is beyond the scope of
this paper. However, we can derive lower and upper limits using a local and a global approximation to the metallicity distribution.

For the former, we follow \cite{LangerNorman06} who computed the formation rate of stars with a metallicity below a threshold
metallicity $Z^{\ast}$ in the local universe. For their fiducial parameters, they found fractions compared to the local total star
formation rate of 0.61, 0.01, 0.0025, and 0.0004 for the four metallicities used in our binary evolution models 
($\zso/4$; $\zso/10$; $\zso/20$; and $\zso/50$; respectively). If the merger delay times were negligible (which they are not; see below), 
and the detection volume were restricted to the low-redshift Universe -- which, from the point of view of the statistics mentioned above 
remains roughly true for redshifts up to $2\sim3$ -- the above factors would need to be applied to obtain local detection rates for the various
metal-poor sources.

A global upper limit to the rates can be obtained by considering the metallicity distribution of all massive stars that have ever formed 
in the Universe. Using the global metallicity distribution provided to us by C.~Kobayashi (based on the simulations presented in \citealt{Taylor15}),
we estimate the fraction of massive stars that have formed with metallicities of $\zso/10$, $\zso/20$ and $\zso/50$ as 0.086, 0.052 and 0.068, 
respectively (using appropriate linear binning).

\subsection{Resulting predictions for aLIGO detection rates}

Using these metallicity weightings, we can estimate ranges for the
aLIGO detection rates (at the design sensitivity) of $19-550\,{\rm
  yr}^{-1}$ for BH+BH mergers below the PISN gap and of $2.1-370\,{\rm
  yr}^{-1}$ above the PISN gap, cf. last column in
Table~\ref{table:rates}. Here, the first quoted number of the ranges
corresponds to the local approximation, the second one to the global
approximation.  Even for the ongoing first science run (O1) of aLIGO
the prospects for detection should be promising. Given that the
sensitivity of aLIGO is currently about 1/3 of the design sensitivity,
the expected detection rate is $\sim\!4\,\%$ of our calculated values
(see last column of Table~\ref{table:rates}).

As the lower-mass BH+BH mergers are more likely to sample the
low-redshift Universe, the lower limit may be more applicable for the
mergers below the PISN gap. On the other hand, as the most massive
BH+BH mergers can be detected throughout most of the visible Universe,
the upper limit may be more appropriate for the mergers above the PISN
gap. Note that for those, the redshift factor $1/(1+z)$ has not yet
been taken into account, as we did not compute the redshift
distribution of events.  In any case, even our lower limits suggest
that aLIGO should detect BH+BH mergers from the MOB scenario. In fact,
the most massive mergers, which probe a large fraction of
the whole Universe, could well be the dominant source for aLIGO
detections {\citep{FlanaganHughes1998,Abadie10}}.

Another factor that helps the detection of BH+BH mergers from
low-metallicity populations, that is not taken into account in the
above estimates, is that, because of the possibly long merger delay
times (see Figure~\ref{fig:delay}), even systems which were formed in the early Universe may merge at low redshift (cf.,
\citealt{MandelMink16}).

A remaining caveat of concern for our estimated detection rates is
related to the relatively low gravitational-wave frequencies of the
more massive BH+BH binaries. The emitted frequencies during in-spiral
are expected to peak approximately at the innermost stable circular
orbit (ISCO) before the plunge-in phase and the actual
merging. However, even determining the ISCO for a merging binary
system is non-trivial and depends, for example, on the spins of the
BHs \citep{balmelli+damour2015}, which requires numerical or
sophisticated (semi-)analytical calculations within general relativity
and cannot simply be estimated using a test particle in a Kerr field.
For the BH+BH mergers above the PISN gap, the emitted frequencies are
most likely $\le 100\,{\rm Hz}$, and with redshift corrections the
frequencies to be detected are easily smaller by a factor of two or
more. Such a low frequency is close to the (seismic noise) edge of the
detection window of aLIGO. However, the waveform amplitudes of the
more massive BH+BH binaries are larger (for a given distance) and are
also enlarged further by a factor of $(1+z)$.  Finally, it may be
possible that higher frequency signals from the ringdown of the
single, rapidly spinning BH produced could be detectable, despite
their expected smaller wave amplitudes.

An important question to address is whether the first generation of
LIGO should have detected such massive BH+BH merger events. Given that
the sensitivity of the first generation of LIGO was about 10 times
lower, the number of detections should have been 1000 times lower.
Therefore even for our upper limits, it is not surprising that there
have been no detections during the previous science runs of the first
generation LIGO detectors \citep{Abadie12}.

\begin{table*}
	\center
	\caption{Fraction of systems per SN that result in double BHs that would merge in less than $13.8\,\mathrm{Gyr}$ (upper 2 rows),
		and aLIGO detection rates (lower 2 rows), assuming that all galaxies have the corresponding metallicity (columns 2--5)
		or are distributed according to our applied integrated metallicity distributions (last column `Integrated Z').
                Here, the first number of the quoted range corresponds to our local approximation, the second one to the
                global approximation, which form lower and upper limits (see text).
		Numbers are given for merging BH+BH binaries both below and above the PISN gap.
		The uncertainties in aLIGO detections rates are mainly caused by mapping the galactic metallicity distribution 
		throughout the Universe (Sect.~\ref{subsubsec:mapping}).}
	\label{table:rates}
	\begin{tabular}{rccccr}
		\hline
		\noalign{\smallskip}
		Metallicity $\longrightarrow$            & $Z_{\odot}/50$      & $Z_{\odot}/20$      & $Z_{\odot}/10$      & $Z_{\odot}/4$       & Integrated Z \\
		\hline
		\noalign{\smallskip}
		$N_{\rm BHBH}/N_{\rm SN}$ below PISN gap & $6.7\times 10^{-4}$ & $1.3\times 10^{-3}$ & $3.4\times 10^{-4}$ & 0                   & $(0.69-13)\times 10^{-5}$ \\ 
		$N_{\rm BHBH}/N_{\rm SN}$ above PISN gap & $2.7\times 10^{-4}$ & 0                   & 0                   & 0                   & $(0.011-1.8)\times 10^{-5}$ \\ 
		\noalign{\smallskip}
		\hline
		\noalign{\smallskip}
		aLIGO rate (yr$^{-1}$) below PISN gap    & 3539                & 5151                & 501                 & 0                   & 19--550 \\ 
		aLIGO rate (yr$^{-1}$) above PISN gap    & 5431                & 0                   & 0                   & 0                   & 2.1--370 \\ 
		\hline
		\noalign{\smallskip}
	\end{tabular}

	\vspace{0.7cm}
	\label{table:rates}
\end{table*}

\section{Concluding remarks}
\label{sect:finish}

We emphasize that -- unlike other channels -- the MOB channel for the
formation of merging BH binaries is quite robust, relying on
reasonably well understood stellar-evolution physics; the main
uncertainty is the treatment of the mixing in rapidly rotating stars,
but even here we can derive some confidence from the fact that our
models are able to reproduce {observed} local counterparts of various stages in
the MOB scenario (see Figure~\ref{fig:cartoon}), such as HD\,5980,
IC10\,X-1 and NGC\,300\,X-1. The MOB channel predicts the formation of
very massive compact BH+BH binaries with a BH mass-ratio close to 1 and a
bimodal BH-mass distribution from BHs formed below and above the
PISN regime.

The detection of GWs from BH+BH mergers with LIGO (and potentially
other current/future GW detectors) will not only start a revolution in
observational astronomy, testing general relativity in its highly dynamic strong-field regime, but will also have a major impact on our understanding of very
massive stars throughout the Universe, including their fate as
gamma-ray bursts or pair-instability supernovae.

\begin{acknowledgements}
PM and NL are grateful to Bill Paxton for his continuous help in
extending the MESA code to contain all the physics required for this
project over the last years. We are thankful to Selma de Mink and Norbert Wex for
helpful discussions, {and to Ed van den Heuvel for useful comments
on an earlier version of this paper}.  We thank Chiaki Kobayashi for providing us with
the global metallicity distribution of massive stars and Andy Fruchter
for plots of the mean metallicity evolution with redshift. 
NL's Alexander von Humboldt Professorship and PP's Humboldt
Research Award provided essential support for this research.
The Geryon2
cluster housed at the Centro de Astro-Ingenieria UC was used for the
calculations performed in this paper. The BASAL PFB-06 CATA, Anillo
ACT-86, FONDEQUIP AIC-57, and QUIMAL 130008 provided funding for
several improvements to the Geryon2 cluster.	
\end{acknowledgements}

\bibliographystyle{aa}

\Online

\appendix

%
%
%

\section{Dynamical implications of black-hole kicks}\label{sec:BHkick}
\begin{figure*}
	\begin{center}
		\includegraphics[width=2.\columnwidth, angle=0]{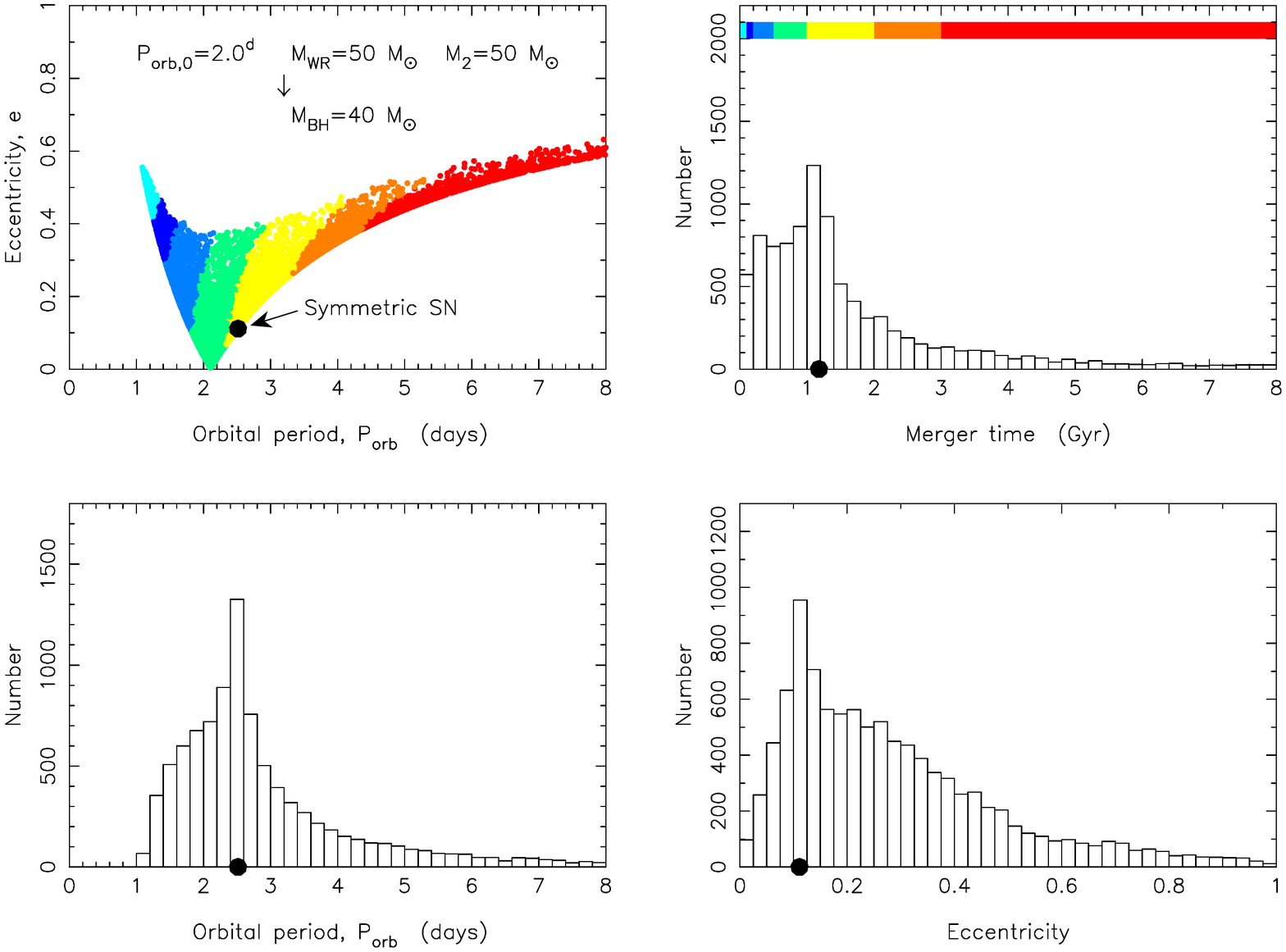}
	\end{center}
	\caption{
		Simulations of the
		dynamical effects of 10\,000 asymmetric SNe on BH binary
		systems.  The initial pre-SN system contains two
		$50\,\mso$ stars in a binary with an orbital period of
		$2.0\,{\rm days}$.  During the formation of a BH, it is
		assumed that $10\,\mso$ is lost instantaneously and a
		kick velocity is imparted with a magnitude between
		$0-300\,{\rm km\,s}^{-1}$, drawn from a flat distribution
		and an isotropic (random) distribution of kick directions.
		The upper-left panel shows the resulting post-SN systems in
		the orbital period--eccentricity plane.  The colours
		indicate the merger time of the post-SN system as a result
		of gravitational-wave radiation (see the distribution in
		the upper-right panel).  The lower panels show the distributions
		in orbital period and eccentricity of the post-SN binaries.
		The case of a purely symmetric SN (no kick) is marked with a
		black dot in all panels.}
	\label{fig:BHkick}
\end{figure*}

Once a tight binary system with two BHs is formed, the
continuous radiation of gravitational waves will give rise to loss of
orbital energy and a resulting shrinkage of the orbit~\citep{Einstein18}. 
The timescale until the two binary components
finally merge depends on the BH masses, the orbital period and the
eccentricity of the system \citep{Peters64}.  The initial parameters
of BH+BH binaries depend not only on the stellar evolution of the
progenitor stars, but also on the physics of the BH-formation process
itself.

The formation of BHs may be accompanied by a momentum kick, similar in
nature to those imparted onto newborn NSs \citep{Janka12}, in
particular if the core of the progenitor does not collapse directly to
a BH but in a two-step process \citep{Brandt95}.  However, whereas
the magnitude of such kicks for NSs is fairly well constrained from
pulsar observations \citep{Hobbs+05}, the magnitude of kicks
imparted onto a BH during its formation is rather uncertain. So far,
the measured masses of observed stellar-mass BHs are all relatively
small: $4-16\,\mso$ for all the Galactic sources
\citep{McClintock+14}, $\sim\!16\,\mso$ for M33~X-7
\citep{Orosz+07} and $24-33\,\mso$ for IC~10~X1
\citep{Prestwich+07}. Their inferred BH kicks from observations and
theoretical arguments span from basically no kicks
\citep{Nelemans+99} to BH kicks of several 100 km/s
\citep{Janka13} based on hydrodynamical kicks associated with
asymmetric mass ejection and subsequent BH acceleration. 

For massive stellar-mass BHs, the situation is different since their
progenitor stars more likely collapse directly to form a BH without a
SN explosion, leading to BHs with very low kick velocities.  Thus, a
bimodality of the BH kick velocity distribution seems possible
\citep{Janka13}.  Another, and possibly related, issue is the amount
of mass loss during BH formation (ejected baryonic mass and rest mass
energy carried away by neutrinos) which also affects the orbital
period and the eccentricity of the post-collapse system.

To quantify the combined effects of mass loss and kicks, we assume in
the following example that a $50\,\mso$ Wolf-Rayet (WR) star 
collapses to form a BH with a gravitational mass of $40\,\mso$
i.e. we assume that $10\,M_\odot$ of mass is lost by a combination of
baryonic mass loss and/or losses through neutrinos from outside 
the event horizon. A large amount of baryonic mass loss may be expected
if the progenitor core is rapidly rotating and the collapse is associated
with a LGRB and an LGRB supernova.

We first consider the no-kick case for a circular pre-collapse system
with two $50\,\mso$ stars and an orbital period of $2.0\,{\rm
	days}$.  If there is no mass loss and no kick imparted to the
newborn BH, the orbital parameters of the binary system remain
unchanged, and the system would merge in $550\,{\rm Myr}$.  The result of an instantaneous mass loss of
$10\,\mso$ during the BH formation produces a post-collapse
system with an orbital period of 2.52~days and an eccentricity of 0.11
(and stellar masses of $40\,\mso$ and $50\,\mso$,
respectively), which will merge in $1180\,{\rm Myr}$. If BH formation is accompanied by an
additional momentum kick, however, the outcome can change
significantly.

In Figure~\ref{fig:BHkick}, we have plotted the dynamical consequences
for a surviving binary system, using the same initial parameters as
before, in which a BH is produced with an asymmetric kick velocity and
instantaneous mass loss, following the recipe of
\cite{Hills83}.  In a BH+BH binary, two kicks may be imparted,
but here we restrict our example to just one kick in order to better
illustrate the principal dynamical effects -- the analysis can easily
be generalized to two kicks without changing the main conclusions. We
performed 10\,000 trials, assuming a flat distribution of the BH kick
magnitudes between $0-300\,{\rm km\,s}^{-1}$ and an isotropic (random)
distribution of directions.  If a randomly orientated kick of a fixed
magnitude of $300\,{\rm km\,s}^{-1}$ ($600\,{\rm km\,s}^{-1}$) were
applied in all cases, it would result in the disruption of about
7.3~per~cent (43~per~cent) of the cases. Using a flat distribution
between $0-300\,{\rm km\,s}^{-1}$, only 0.4~per~cent of all systems
are disrupted.  Thus even for a wide range of assumed BH kick values,
the survival rate of BH+BH binaries only changes with less than a
factor of two.

More important for our investigation here is the merger timescale due
to gravitational-wave radiation.  Figure~\ref{fig:BHkick} shows that the
effect of a kick can either widen or shorten the post-collapse orbit.
However, the merger timescale for a binary with given component masses
is a function of both orbital period and eccentricity. Given that
systems which widen more during BH formation (as a consequence of
instantaneous effective mass loss and a kick) are also the systems
attaining the larger eccentricities, the net effect of a kick on the
resulting merger time is surprising small.  For the case of BH
formation with no kick, the merger timescale of the resulting binary
is 1180~Myr.  Using a flat kick distribution between
$0-300\,{\rm km\,s}^{-1}$ results in roughly half (47~per~cent) of the
surviving systems to merge on a shorter timescale, and the other half
to merge on a longer timescale (and only 2.5~per~cent exceeding a
Hubble time).

Applying a strong kick of $600\,{\rm km\,s}^{-1}$ actually causes a
larger fraction of surviving systems to merge (67~per~cent) on a shorter
timescale ($<1180\,{\rm Myr}$) compared to the symmetric case with no
kick. Therefore, we can safely conclude that although BH kicks may, in
general, widen a number of systems, the resulting merger timescale
distribution will not change one of the main findings of this paper,
namely that the LIGO detection rate is likely to be dominated by quite
massive BH+BH mergers.

\section{Comparison to the standard BH+BH formation scenario}\label{sec:BHstd}

\begin{figure}
	\begin{center}
		\includegraphics[width=1.0\columnwidth, angle=0]{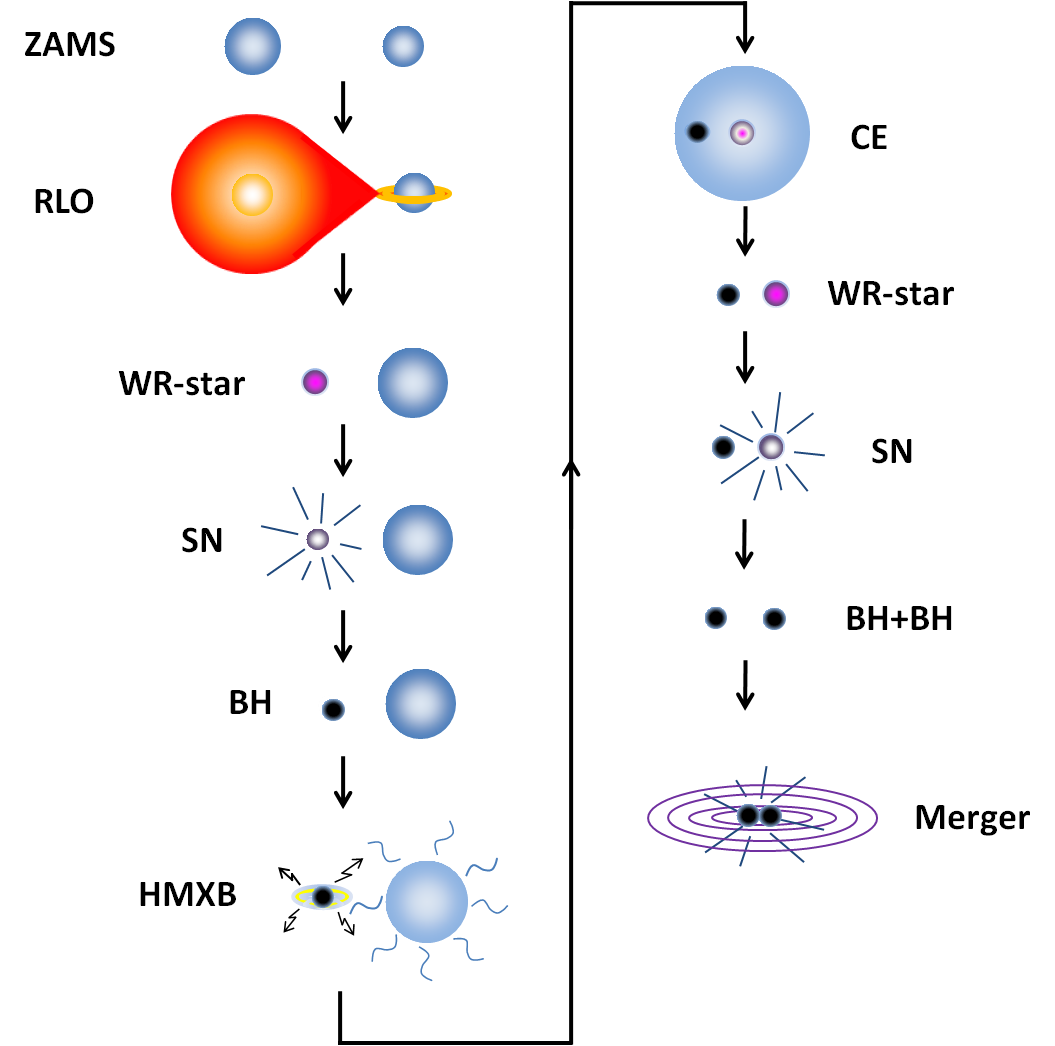}
	\end{center}
	\caption{ Illustration of the binary stellar evolution leading to
		a BH+BH binary according to the standard scenario.  Acronyms
		used in the figure. ZAMS: zero-age main sequence; RLO:
		Roche-lobe overflow (mass transfer); WR-star: Wolf-Rayet
		star; SN: supernova; BH: black hole; HMXB: high-mass X-ray
		binary; CE: common envelope.} 
	\label{fig:BHstd}
\end{figure}

The standard formation scenario of BH+BH binaries involves a number of
highly uncertain aspects of binary interactions
(Figure~\ref{fig:BHstd}, {\citealt{Tauris2006}}). The main uncertainties include, in particular,
the treatment of common-envelope (CE) evolution \citep{Ivanova13}
and the efficiency of accretion and spin-up from mass transfer.  These
lead to uncertainties in the expected merger rates of several orders
of magnitude \citep{Abadie10}. In contrast, the MOB scenario presented here
relies mostly on reasonably well understood physics of the evolution
of massive stars, although there are still significant uncertainties
in the treatment of, e.g., stellar winds \citep{Langer12},
rotational mixing \citep{MaederMeynet12} and the BH formation itself
\citep{Heger+13,Ugliano+12,PejchaThompsom15}.

To produce a tight BH binary that will merge within a Hubble time, one
could consider a very close massive binary system with an initial
orbital period of a few days.  The problem with such a model is that
such systems would mostly be expected to merge early during their
evolution when the more massive star evolves off the main sequence,
expands and starts to transfer mass to its companion star. Although
other models have been proposed which evolve without a CE phase,
e.g. to explain the formation of IC~10~X-1 \citep{Mink09} and
M33~X-7 \citep{Valsecchi+10}, they often require some degree of
fine-tuning to work and did not follow the evolution to the end to
produce a binary with two BHs.

To avoid this problem, it has been common practice to model the
formation of BH+BH systems starting from relatively wide systems and
let the systems evolve through a CE phase following the high-mass
X-ray binary (HMXB) phase after the formation of the first BH (see
Figure~\ref{fig:BHstd}). There are presently no self-consistent
hydrodynamical simulations for modelling the spiral-in of BHs inside a
massive envelope; in particular, it is unclear whether the BH will
experience hypercritical accretion, and under what conditions the
systems will merge completely; all of this leads to large
uncertainties in the number of post-CE systems and their separations
\citep{Ivanova13}.

Another problem that is often ignored is the fact that the fate of a
massive star in a binary depends on when it loses its hydrogen-rich
envelope. As first pointed out by Brown \citep{Brown99} and confirmed in later
calculations \citep{Brown01} (see also Petermann, Langer \& Podsiadlowski;
in preparation), if a massive star loses its hydrogen-rich envelope before or
early during helium core burning, it ends its evolution with a much
smaller iron core and is more likely to produce a NS than a
BH.  This means that the formation of a BH in a close binary
may require that its progenitor loses its envelope only after helium
core burning.  In order to produce not just one but two BHs,
this may require extreme fine-tuning and may even be impossible in the
standard scenario in Figure~\ref{fig:BHstd}. The problem may be avoided
in the so-called double-core scenario \citep{Brown95,Dewi06} where the BH
progenitors both evolve beyond helium core burning before experiencing a
CE phase in which the cores of both stars spiral in, producing a close
binary of two helium stars that subsequently collapse to form BHs. However, this scenario requires significant fine-tuning since
the initial masses of the two stars have to be very close to each
other ($q_\mathrm{i}>0.96$), and it may be impossible to produce quite massive BHs as very massive stars tend to avoid mass transfer after helium core burning
(although at sufficiently low metallicity ($Z\simle 0.1\,Z_\odot$)
this may be possible).

Given that all measured stellar-mass BHs in the Milky~Way have masses
in the range of $4-16\,\mso$, most population-synthesis models
used for estimating LIGO detection rates were previously 
restricted to initial progenitor stellar masses up to about
$100\,\mso$.  The discovery of very massive stars
\citep{Crowther10,Hainich+14} in the R136 region of the Large
Magellanic Cloud with masses up to $300\,\mso$, however,
suggests that BH+BH binaries may form with significantly more massive
components, thus enabling LIGO to detect the merger of such massive
BH+BH binaries, with chirp masses easily exceeding $30\,\mso$,
out to long distances (see discussion in section \ref{sect:rates}).

It had previously been argued \citep{Belczynski+14,Rodriguez15}
that massive BH+BH binaries to be potentially detected by LIGO can
only form via dynamical channels in dense stellar environments. In
this letter, we have demonstrated that close binaries in the Galactic
disk with very massive stars undergoing chemically homogeneous evolution 
(and which therefore do
not expand after leaving the main sequence) can form massive BH+BH
binaries that merge within a Hubble time at sufficiently low
metallicity.  An important consequence of our scenario is that
massive BHs of a given mass can be produced from stars with a lower
ZAMS mass -- especially at low metallicity -- compared to the standard
BH+BH formation scenario.

So far, no stellar-mass BH+BH binaries have been discovered anywhere,
and, in the Milky~Way, there is only one known potential progenitor
system, Cyg~X-3. The nature of the compact component in this system is
still uncertain; \cite{Belczynski13} have
argued that it contains a $2-4.5\,\mso$ BH and a
$7.5-14.2\,\mso$ WR-star, making it a potential progenitor for a
BH+BH system. However, the final destiny of this system is unclear, and
it could also become a BH-NS binary or even NS+NS binary (if the
first-formed compact object is a neutron star); it is also possible
that the system is disrupted as a result of the explosion of the WR-star.

\end{document}